\documentclass[12pt]{article}
\usepackage{epsfig}
\usepackage{cite}
\def\DESepsf(#1 width #2){\epsfxsize=#2 \epsfbox{#1}}
\begin{document}

\title{\Large\bf See-saw Fermion Masses in an SO(10) GUT}

\author{Kaushik Bhattacharya ${}^1$ \footnote{\large\, kaushik@prl.res.in} ,
C.R. Das ${}^2$ \footnote{\large\, crdas@imsc.res.in} , 
Bipin R. Desai ${}^3$ \footnote{\large\, bipin.desai@ucr.edu, bipin@earthlink.net} ,\\
G. Rajasekaran ${}^2$  \footnote{\large\, graj@imsc.res.in}\,\, and 
Utpal Sarkar ${}^{1}$  \footnote{\large\, utpal@prl.res.in}\\ \\
{\sl ${}^1$ Physical Research Laboratory, Ahmedabad 380 009, India}\\
{\sl ${}^2$ Institute of Mathematical Sciences, Chennai 600 113, India}\\
{\sl ${}^3$ Physics Department, University of California, }\\
{\sl Riverside, CA 92521, USA}}

\date{}
\maketitle

\begin{abstract}

\noindent 
In this work we study an SO(10) GUT model with minimum Higgs
representations belonging only to the {\bf 210} and {\bf 16} dimensional
representations of SO(10). We add a singlet fermion $S$ in
addition to the usual {\bf 16} dimensional representation containing
quarks and leptons. There are no Higgs bi-doublets and so charged
fermion masses come from one-loop corrections. Consequently all the fermion
masses, Dirac and Majorana, are of the see-saw type.  We minimize the
Higgs potential and show how the left-right symmetry is broken in our
model where it is assumed that a $D$-parity odd Higgs field gets a
vacuum expectation value at the grand unification scale. From the
renormalization group equations we infer that in our model unification
happens at $10^{15}$ GeV and left-right symmetry can be
extended up to some values just above $10^{11}$ GeV. The Yukawa
sector of our model is completely different from most of the standard
grand unified theories and we explicitly show how the Yukawa sector
will look like in the different phases and briefly comment on the
running of the top quark mass. We end with a brief analysis of lepton
number asymmetry generated from the interactions in our model.

\end{abstract}

\clearpage\newpage

\parindent=1cm
\baselineskip 18pt

\section{Introduction}
The SO(10) grand unified theory has several interesting features
\cite{min1,min2,min3,min4,min5}.  It can accommodate left-right
symmetry as one of the intermediate symmetry and hence provides an
explanation of parity violation \cite{lr}.  $(B-L)$ is a generator of the group
SO(10) and hence lepton number violation takes place
spontaneously. This explains the origin of lepton number violation and
neutrino Majorana mass naturally. The smallness of neutrino masses is
assured by the see-saw mechanism, so that by keeping the scale of
$(B-L)$ violation high the smallness of the neutrino mass is
guaranteed. Quarks and leptons are treated equally in SO(10)
GUT. Gauge coupling unification is consistent with all low energy
results.

On the one hand there are many attractive features of the SO(10)
GUT, on the other the predictability becomes low. Depending on the
symmetry breaking pattern and Higgs scalar contents, the model can
have widely differing predictions. Attempts have been made to construct
a minimal model. In one approach the minimal model is
constructed with minimum numbers of parameters, while in the other
approach minimum numbers of Higgs scalars are included in the
model. There are also models without any intention of minimality
or simplicity, where the main aim is to explain all experiments
and have maximum predictability.

In the present article we shall study an SO(10) GUT, which has the
minimum dimensions of the Higgs scalars. In any SO(10) GUT the minimal
number of Higgs scalar includes an ${\rm SU(2)}_L$ symmetry breaking Higgs
scalar, which can give masses to the fermions and some scalars that
can break the $(B-L)$ symmetry along with the ${\rm SU(2)}_R$ symmetry
and can give Majorana masses to the neutrinos.  In addition, there is
one Higgs scalar which breaks the group SO(10) GUT. One conventional
model includes a Higgs bi-doublet (a {\bf 10}-plet of SO(10) Higgs
scalar, which is doublet under both the ${\rm SU(2)}_L$ and ${\rm
SU(2)}_R$ groups) and both ${\rm SU(2)}_L$ and ${\rm SU(2)}_R$ Higgs
triplets. In such models with triplet Higgs scalars neutrinos acquire
masses at the tree level. The SO(10) representation that contains this
Higgs scalar is of {\bf 126} dimensions. In another version of the model,
one breaks the left-right symmetry and the $(B-L)$ symmetry by
doublets of ${\rm SU(2)}_L$ and ${\rm SU(2)}_R$ groups. 
This Higgs field belongs to
a {\bf 16}-plet representation of SO(10). For symmetry breaking and giving
fermion masses another {\bf 10}-plet Higgs scalar is introduced, which
contains the bi-doublet Higgs and hence gives tree level masses to the
fermions and the neutrinos acquire only Dirac masses of the order
of other fermion masses. There is no Majorana mass term for the
neutrinos and hence see-saw mechanism is not possible. However,
there exist effective higher-dimensional operators, 
which can give correct Majorana 
masses to both the left-handed and right-handed neutrinos. 

Recently it has been pointed out that it is possible to consider an
SO(10) GUT, which does not have any Higgs bi-doublet scalar belonging
to a {\bf 10}-plet of SO(10) \cite{Desai:2004xf}. The Higgs scalar
that breaks left-right symmetry and the $(B-L)$ symmetry belongs to a
{\bf 16}-plet of Higgs scalar. Since tree level fermion masses are not
allowed without the bi-doublet scalar, all fermion masses come from
higher-dimensional operators in the see-saw form.  In supersymmetric
theories the non-renormalizablity theorem does not allow radiative
generation of such higher dimensional operators. We shall then
restrict ourselves to only non-supersymmetric SO(10) GUT.  The source
of see-saw suppression for the fermion number violating Majorana mass
terms are different from the source of see-saw suppression for the
fermion number conserving Dirac mass terms, which maintain the large
hierarchy between the charged fermion masses and the neutrino
masses. In this article we shall study some aspects of this model in
detail.

In the next section we shall discuss the model. In 
sec. \ref{hpot} we shall present details of the scalar potential
minimization and the allowed symmetry breaking pattern and in sec. \ref{fmass}
the generation of fermion masses is discussed. We shall then
study the renormalization group equation for this model with
the specific choice of the Higgs scalars. In sec. \ref{gcu} we shall study the
gauge coupling unification and in sec. \ref{yukawa} the Yukawa coupling 
evolution for the different fermions. Since all fermion masses have the same
see-saw origin, the perturbative unification becomes an important
question in these models. Some of the Yukawa couplings could
become large, although the effective fermion masses still remains
small. In sec. \ref{lepto} leptogenesis in our model is discussed and in the 
last section we summarize our results.
\section{The Model}
\label{sbpat}
The starting point for any SO(10) GUT is the choice of the symmetry
breaking pattern. There exists many chains of symmetry breaking
pattern, which are all consistent with our present knowledge. So
the particular choice of a symmetry breaking pattern 
defines a specific model. We shall consider a symmetry breaking
pattern, which requires a minimum number of Higgs scalars, given
by:
\begin{eqnarray}
{\rm SO(10)}&{\stackrel{\Phi_D}\longrightarrow}&{\rm SU(4)}_C
\times{\rm SU(2)}_L\times{\rm SU(2)}_R \quad[{\rm G}_{422}]
\nonumber \\
&{\stackrel{\Phi_4} \longrightarrow}&
{\rm SU(3)}_C\times{\rm SU(2)}_L\times{\rm SU(2)}_R\times{\rm U(1)}_{B-L}
\quad [{\rm G}_{3221}] \nonumber\\
&{\stackrel{\chi_R}\longrightarrow}& {\rm
SU(3)}_C\times{\rm SU(2)}_L\times{\rm U(1)}_Y \quad[{\rm G}_{321}] \nonumber \\
&{\stackrel{\chi_L}\longrightarrow}&{\rm SU(3)}_C\times {\rm
U(1)}_{\rm em}\,, \label{sbp}
\end{eqnarray}
where the Higgs fields responsible for the symmetry breakings,
$\Phi_D$, $\Phi_4$, $\chi_R$ and $\chi_L$ are explicitly shown in
the above equation. Both $\Phi_D$ and $\Phi_4$ are contained in
$\Phi$, which transforms as the {\bf 210} dimensional
representation  of SO(10). The {\bf 210}-plet decomposes under the
Pati-Salam subgroup ${\rm G}_{422}$ of SO(10) as,
\begin{eqnarray}
\Phi\equiv{\bf 210}&=&(\bf{1,1,1}) + (\bf{6,2,2}) + (\bf{15,3,1}) + (\bf{15,1,3})\nonumber\\
&& +\, (\bf{15,1,1}) + (\bf{10,2,2}) + (\overline{\bf 10}, \bf{2,2})\,.
\end{eqnarray}
In the above decomposition $\Phi_D$ corresponds to $({\bf 1,1,1})$ and
$\Phi_4$ corresponds to ({\bf 15,1,1}).

The Higgs fields $\chi_L$ and $\chi^*_R$ belong to the {\bf 16}
dimensional spinor representation $\Gamma$ of SO(10) and the other
fields $\chi_R$ and $\chi^*_L$ belong to the conjugate
representation $\overline{\bf 16}$, called $\Gamma^\dagger$, of
SO(10). The group transformation properties of the $\chi$ fields under
${\rm G}_{3221}$ and ${\rm G}_{422}$ are as follows:
\begin{eqnarray}
\begin{array}{ccccc}
\chi_L &\equiv& ({\bf 1,2,1},-1)\subset({\bf 4,2,1})\subset{\bf 16} &\equiv& \Gamma\,,\\
\chi^*_{R} &\equiv& ({\bf 1,1,2},1)\subset({\overline {\bf 4}},{\bf 1,2})
\subset {\bf 16} &\equiv&
\Gamma\,,\\
\chi_R &\equiv& ({\bf 1,1,2},-1)\subset({\bf
4,1,2})\subset{\overline{\bf 16}} &\equiv& {\Gamma^\dagger}\,,\\
\chi^*_{L} &\equiv& ({\bf 1,2,1},1)\subset({\overline {\bf 4}},{\bf 2,1})\subset{\overline {\bf 16}}
&\equiv& {\Gamma^\dagger}\,.
\end{array}
\label{higgsconj}
\end{eqnarray}
In the above equations $({\bf x,y,z},w)$, and $({\bf x,y,z})$
denote the group transformation properties of the Higgs fields
under ${\rm G}_{3221}$ and ${\rm G}_{422}$.

At this stage we shall digress to discuss one important feature of the left-right
symmetric models, namely the question of parity ${\cal P}$.  The
discrete $Z_2$ symmetry, that interchanges the two SU(2) subgroups of
the Lorentz group O(3,1), is called the parity.  This parity can
be identified as the discrete $Z_2$ symmetry operator that
interchanges the groups ${\rm SU(2)}_L$ and ${\rm SU(2)}_R$ of the left-right
symmetric model, which implies that under parity $W_L \stackrel{\cal
P}{\longleftrightarrow} W_R$. This definition extends to scalars
also. That is, an ${\rm SU(2)}_L$ doublet scalar field $\chi_L$ will
transform to an ${\rm SU(2)}_R$ doublet scalar field $\chi_R$ under the
operation of parity $\chi_L \stackrel{\cal P}{\longleftrightarrow}
\chi_R$.  In the conventional left-right symmetric models, the parity
is spontaneously broken along with the group ${\rm SU(2)}_R$. In other
words, when the left-right symmetric group ${\rm SU(2)}_L \times {\rm SU(2)}_R$ is
spontaneously broken, parity is also spontaneously broken.

There is another possibility of breaking parity spontaneously
without breaking the left-right symmetric group. Since the
scalar fields transform trivially under the Lorentz group,
the VEV of a parity odd field can break parity spontaneously
without breaking the left-right symmetry. Unlike the conventional
case, now the parity acting on the fermions and vector bosons
is not spontaneously broken. To distinguish these two cases,
this second type of parity is called a $D$-parity. Thus
when $D$-parity is broken, the left-handed and right-handed
scalars can have different mass and VEV and hence the 
gauge coupling constants of ${\rm SU(2)}_L$ and ${\rm SU(2)}_R$ can also
be different. In the present model $D$-parity plays a very
crucial role, both for symmetry breaking as well as for 
fermion masses. It also plays some role in gauge coupling 
unification. 

The ${\bf 210}$ representation of SO(10) is a totally antisymmetric
tensor of rank four $\Phi_{abcd}$ and the singlet $\Phi_D$ is the
component $\Phi_{6789}$ in the notation, in which, $a,b,c,d =
0,1,..,5$ are ${\rm SO(6)}$ indices and $a,b,c,d = 6,7,8,9$ are ${\rm SO(4)}$
indices. Thus under $D$-parity $\Phi_D$ is odd ($\Phi_D \to - \Phi_D$)
and consequently when it gets its vacuum expectation value (VEV) at
the GUT scale, $M_U$, it breaks the left-right parity of the
theory. Due to this spontaneous breaking of the left-right parity at
the $M_U$ scale we will have $\langle \chi_L \rangle \neq \langle
\chi_R \rangle$ at a lower energy scale. This $D$-parity odd field is
also required to give masses to the light neutrinos.

Next we write down the fermions in our model and their group
transformation properties. The left-handed quarks, leptons,
anti-quarks and anti-leptons belong to a {\bf 16}-plet representation of
SO(10), which transform under ${\rm G}_{422}$ as:
\begin{eqnarray}
\psi_{i\,L}\equiv{\bf 16}=(\bf{4,2,1}) + ({\overline{\bf 4}}, \bf{1, 2})\,,
\label{flso10}
\end{eqnarray}
$i=1, 2, 3$ is the generation index. The right-handed fermions and
anti-fermions belong to the conjugate representation,
\begin{eqnarray}
\psi_{i\,R}\equiv{\overline{\bf 16}}=({\overline{\bf 4}}, \bf{2, 1}) + (\bf{4,1,2})\,.
\label{frso10}
\end{eqnarray}
In addition to the above mentioned conventional particles our
model consists of an extra SO(10) gauge singlet fermion per
generation:
\begin{eqnarray}
S_{iL}\equiv{\bf 1}=({\bf 1,1,1})\,, \label{sing}
\end{eqnarray}
$i=1,2,3$.

Under ${\rm G}_{3221}$ the states $(\bf{4,2,1})$ and
$({\overline{\bf 4}}, \bf{1, 2})$ transform as:
\begin{eqnarray}
(\bf{4,2,1})&=&({\bf 3,2,1},\frac{1}{3})+({\bf 1,2,1},-1)\,,\\
({\overline{\bf 4}}, \bf{1, 2})&=&({\overline{\bf 3}},{\bf 1,2},-\frac{1}{3})+({\bf 1,1,2},1)\,,
\end{eqnarray}
and as a result the fermions can be labelled as:
\begin{eqnarray}
q_L&=&
\left(\begin{array}{c}
u\\
d
\end{array}
\right)_L\equiv ({\bf 3,2,1},\frac{1}{3})\,,\\
\ell_L&=&
\left(\begin{array}{c}
\nu\\
e
\end{array}
\right)_L\equiv ({\bf 1,2,1},-1)\,,
\label{lql}
\end{eqnarray}
and
\begin{eqnarray}
q^{\,\,\,\,c}_R=q^c_{\,\,L}&=&
\left(\begin{array}{c}
d^{\,c}\\
u^c
\end{array}
\right)_L\equiv ({\overline{\bf 3}},{\bf 1,2},-\frac{1}{3})\,,\\
\ell^{\,\,\,\,c}_R=\ell^c_{\,\,L}&=&
\left(\begin{array}{c}
e^c\\
\nu^c
\end{array}
\right)_L\equiv ({\bf 1,1,2},1)\,.
\label{rql}
\end{eqnarray}

The generators of the left-right symmetry group ${\rm G}_{3221}$ are related
to the electric charge of the particles by,
\begin{eqnarray}
Q=T_{3L}+T_{3R}+\frac{(B-L)}{2}=T_{3L}+\frac{Y}{2}\,,
\label{charge}
\end{eqnarray}
where
\begin{eqnarray}
Y=T_{3R}+\frac{(B-L)}{2}\,.
\end{eqnarray}
In the conventional left-right symmetric models there is one
bi-doublet Higgs scalar $\phi \equiv ({\bf 1,2,2,}0)$, which gives
masses to quarks and charged leptons and a Dirac mass to the
neutrinos through its couplings of the form $\overline \psi_L
\psi_R \phi$. In addition, there are triplet Higgs scalars
$\Delta_L \equiv ({\bf 1,3,1},-2)$ and $\Delta_R \equiv ({\bf
1,1,3,}-2)$, which can give Majorana masses to the left-handed and
right-handed neutrinos through the couplings $\psi_{L,R}
\psi_{L,R} \Delta_{L,R}$. In our present model all these Higgs
scalars $\phi$, $\Delta_L$ and $\Delta_R$ are absent and hence
there are no tree level fermion masses for the quarks and the
leptons. After discussing the structure of the Higgs vacuum
expectation values in this model, we shall come back to the
question of fermion masses.
\section{Minimization of the scalar potential and left-right symmetry breaking}
\label{hpot}
In the conventional left-right symmetric models, the combinations
of the Higgs fields, $\phi$, $\Delta_L$ and $\Delta_R$ ensures
that for certain choices of parameters, $\Delta_R$ can acquire a
very large VEV compared to other fields breaking left-right
symmetry at a large scale. It is clear that in the absence of the
field $\phi$, both the fields $\Delta_{L,R}$ would acquire equal
VEVs. It has also been shown that in the absence of the field
$\phi$ in a left-right symmetric model with only the doublet Higgs
scalars $\chi_L$ and $\chi_R$, the minimization of the potential
would result in equal VEVs for both $\chi_L$ and $\chi_R$, which
would lead to inconsistency and parity will be conserved at low
energy. This problem could be solved if parity is broken in these
theories either explicitly or spontaneously.

In the present model this problem does not occur. It was mentioned
in the original version of the model that the $D$-parity odd singlet
field $\Phi_D \equiv {\bf (1,1,1)}$ under G${}_{422}$ contained in
the ${\bf 210}$ representation would allow $\langle \chi_R \rangle
\gg \langle \chi_L \rangle$ and break left-right symmetry at some
high scale. In this section we shall minimize the scalar potential
and discuss the various possible solutions, which allows
left-right symmetry breaking at some high scale.

Let us consider the Higgs potential \cite{Desai:2004xf}:
\begin{eqnarray}
{\cal L}_s &=& m^2_\Phi \Phi^2 + a \Phi^3 +
\frac{\lambda_\Phi}{4!}\Phi^4 + m^2_\Gamma (\Gamma^\dagger \Gamma)
+ \frac{\lambda_\Gamma}{4}(\Gamma^\dagger \Gamma)^2\nonumber\\
&&+\,\frac{\lambda'_\Gamma}{4}[\Gamma^4 + (\Gamma^\dagger)^4]
+ M_D \Phi (\Gamma^\dagger \Gamma) + \lambda_{\Phi \Gamma}
\Phi^2 (\Gamma^\dagger \Gamma)\,.
\label{hp}
\end{eqnarray}
The coupling $\Phi(\Gamma^\dagger \Gamma)$ is the most important
term that is required for the left-right breaking to take place at
a higher scale compared to standard model symmetry breaking.
$D$-parity is broken when $\Phi_D$ acquires a non-vanishing VEV,
$\langle \Phi_D \rangle = \eta$,  at the $M_U$ scale, since
$\Phi_D$ is odd under $D$-parity. $\Phi_4$ will get a non-vanishing
VEV at the $M_X$ scale. There will be many terms in Eq.~(\ref{hp})
including $\langle \Phi_4 \rangle$, but as we are analyzing the
structure of Eq.~(\ref{hp}) in the ${\rm G}_{3221}$ phase and
mainly interested on the VEVs of the $\chi$ fields, we do not
explicitly write down the terms including $\langle \Phi_4
\rangle$. $\langle \Phi_4 \rangle$ has no important contribution
in the expressions of the VEVs of the $\chi$ fields.

We now discuss the masses of the components of $\Gamma$ and the
VEVs. The scalar potential responsible for the masses of the
fields $\chi_L$ and $\chi_R$ is given by,
\begin{eqnarray}
V &=& m_\Gamma^2 ( |\chi_R|^2 + |\chi_L|^2 ) + M_D ~ \eta
(|\chi_R|^2 - |\chi_L|^2 )
\nonumber \\
&&+\, \lambda_{\Phi \Gamma} ~\eta^2 (|\chi_R|^2 + |\chi_L|^2 )\,.
\end{eqnarray}
The masses of these fields are then given by,
\begin{eqnarray}
\mu_L^2 &=& m_\Gamma^2 - M_D ~\eta + \lambda_{\Phi \Gamma}
 ~\eta^2\,,\nonumber \\
\mu_R^2 &=& m_\Gamma^2 + M_D ~\eta + \lambda_{\Phi \Gamma}
 ~\eta^2\,.
\end{eqnarray}
If $D$-parity is conserved, $\eta = 0$ and the masses of both
$\chi_L$ and $\chi_R$ become equal. Since the VEV of $\Phi_D$
breaks $D$-parity, it will be possible to fine tune parameters to
obtain the mass of $\chi_L$ to be orders of magnitude smaller than
the mass of $\chi_R$. From phenomenological consideration we also
require
\begin{eqnarray}
\langle \chi_L \rangle &=& v_L \sim \mu_L \sim 100~{\rm GeV}\,,
\nonumber \\
\langle \chi_R \rangle &=& v_R \sim \mu_R \sim M_U \gg v_L\,.
\end{eqnarray}
We shall next check if this widely different VEVs for $\chi_L$
and $\chi_R$ is possible. $v_L$ breaks the electroweak symmetry,
while $v_R$ breaks the left-right symmetry at a very high scale,
close to the GUT scale.

We denote the VEVs of the fields $\chi_L$ and $\chi_R$ as:
\begin{eqnarray}
\langle \chi_L \rangle =
\left(\begin{array}{c}
v_L\\
0
\end{array}
\right),\,\,
\langle \chi^*_R \rangle =
\left(\begin{array}{c}
0\\
v^*_R
\end{array}
\right),\,\,
\langle \chi_R \rangle =
\left(\begin{array}{c}
v_R\\
0
\end{array}
\right),\,\,
\langle \chi^*_L \rangle =
\left(\begin{array}{c}
0\\
v^*_L
\end{array}
\right).
\label{vevs}
\end{eqnarray}
Instead of minimizing the potential, we shall first write down the
potential in terms of the VEVs of the fields and then find the
conditions satisfied by the VEVs. With the above VEVs we can
write the Higgs potential in the ${\rm G}_{3221}$ phase as:
\begin{eqnarray}
V&=&-\,m^2_\Phi(\eta^2 + \cdot \cdot \cdot) - a \,(\eta^3 + \cdot
\cdot \cdot)
- \frac{\lambda_\Phi}{4!}(\eta^4 + \cdot \cdot \cdot)\nonumber\\
&&+\,[m^2_\Gamma + \lambda_{\Phi \Gamma}(\eta^2 +\cdot \cdot \cdot)]
(v^*_R v_R + v^*_L v_L) - \frac{\lambda_\Gamma}{4}(v^*_R v_R +
v^*_L v_L)^2
\nonumber\\
&&-\,\frac{\lambda'_\Gamma}{4}(v^4_L + v^4_R + v^{*\,4}_L +
v^{*\,4}_R) + M_D \eta (v^*_R v_R - v^*_L v_L)\,. \label{pot1}
\end{eqnarray}
The $\cdot \cdot \cdot$ symbols in the above equation stands for
terms containing $\langle \Phi_4 \rangle$.

Setting $v_R=v^*_R$ and $v_L=v^*_L$, which amounts to saying that
there is no CP violation and all VEVs are considered to be real,
the extremum conditions of $V$ comes out to be:
\begin{eqnarray}
\frac{\partial V}{\partial v_L}&=& 2v_L \left[ m^2_\Gamma +
(\lambda_{\Phi \Gamma} \eta^2 +\cdot \cdot \cdot) -
\frac{\lambda_\Gamma}{2} (v^{2}_L + v^{2}_R) -\lambda'_\Gamma
v^2_L - M_D \eta \right] =
0\,,\nonumber \\
\label{vl}\\
\frac{\partial V}{\partial v_R}&=& 2v_R \left[ m^2_\Gamma +
(\lambda_{\Phi \Gamma} \eta^2 +\cdot \cdot \cdot) -
\frac{\lambda_\Gamma}{2} (v^{2}_L + v^{2}_R) -\lambda'_\Gamma
v^2_R + M_D \eta \right] = 0\,.\nonumber \\ \label{vr}
\end{eqnarray}
The above equations imply,
\begin{eqnarray}
v_L\left(\frac{\partial V}{\partial v_R}\right) - v_R
\left(\frac{\partial V}{\partial v_L}\right) = 2v_L
v_R[\lambda'_\Gamma(v^2_L - v^2_R) + 2 M_D\eta]=0\,. \label{vlvr}
\end{eqnarray}
Neglecting the trivial solution $v_L=v_R=0$, the other interesting
relation between $v_L$ and $v_R$ that comes out from the above
equation is,
\begin{eqnarray}
v^2_R - v^2_L = \frac{2 M_D\eta}{\lambda'_\Gamma}\,.
\label{vlvrsq}
\end{eqnarray}
Two things can be noted from the above equation. First as it was
stated previously, in understanding the relation between $v_L$ and
$v_R$ we do not require the VEV of $\Phi_4$. Secondly if $\eta$
has some value comparable to $M_U$ and $\lambda'_\Gamma$ is not
too high, then it is apparent from Eq.~(\ref{vlvrsq}) that $v_R
\gg v_L$. If the energy scale where $\chi_R$ and $\chi^*_R$
gets a non vanishing VEV be $M_R$ then we can say that $M_R \gg
M_W$ where $M_W\simeq 100$ GeV. Thus this model allows left-right
symmetry breaking at a much higher scale compared to the standard
model symmetry breaking scale.

It is clear from the above discussions that this model works only
if $D$-parity is broken spontaneously. In addition, severe fine
tuning is required to obtain and maintain this solution. To make
the masses of $\chi_L$ and $\chi_R$ different $\mu_L \neq \mu_R$,
a fine tuning is required. Then the next fine tuning is required
to keep the VEV $v_L$ to be orders of magnitude smaller than
$v_R$. This is the usual fine tuning required in all
non-supersymmetric theories. We can write Eq.~(\ref{vl}) as
\begin{eqnarray}
\frac{\partial V}{\partial v_L}= v_L \left[ \mu_L^2 -
\frac{\lambda_\Gamma}{2} (v^{2}_L + v^{2}_R) -\lambda'_\Gamma
v^2_L \right] = 0\,.
\end{eqnarray}
Since the VEV $v_L$ will be proportional to $\mu_L$, a fine tuning
is performed to keep $\mu_L \sim 100$ GeV. The second fine tuning
makes sure that the VEV $v_R$ does not destabilize the VEV of
$v_L$ through radiative corrections.

\section{Fermion masses} \label{fmass}
%
In the left-right symmetric theories the left-handed fermions are
doublets under ${\rm SU(2)}_L$ and the right-handed fermions are
doublets under ${\rm SU(2)}_R$. Hence the fermion masses would require
a bi-doublet Higgs scalar. Following our discussions at the end of
sec. \ref{sbpat}, it is clear that in the present model there are
no Yukawa couplings giving Dirac or Majorana masses to the quarks and
leptons. In this model both the Majorana and the Dirac masses
originate from dimension-5 effective operators, given by:
\begin{eqnarray}
\begin{array}{rclcrcl}
{\cal O}_1 &=& \displaystyle{1 \over M_D}(\overline{q_L}
{\chi_L})(q_R \chi^*_{R})\,, & \quad & {\cal O}_2 &=&  \displaystyle{1
\over M_D}(\overline{q_L} \chi^*_{L})(q_R {\chi}_R)\,, \cr &&&&&& \cr
{\cal O}_3 &=&  \displaystyle{1 \over M'_D}(\overline{\ell_L}
{\chi_L})(\ell_R \chi^*_{R})\,, && {\cal O}_4 &=& \displaystyle{1
\over M'_D}(\overline{\ell_L} \chi^*_{L})(\ell_R {\chi}_R)\,, \cr
&&&&&& \cr {\cal O}_5 &=& \displaystyle{1 \over M_M}(\ell_L
\chi^*_{L})(\ell_L \chi^*_{L})\,, && {\cal O}_6 &=& \displaystyle{1
\over M_M}(\ell_R \chi^*_{R})(\ell_R \chi^*_{R})\,,
\end{array}
\label{opprod}
\end{eqnarray}
where $M_D$, $M'_D$  and $M_M$ are some heavy mass scales in the theory.  In
general, the mass scales appearing in the operators which contribute
to the Dirac masses ($M_D$, $M'_D$) and the mass scales that appear in the
operators contributing to the Majorana masses ($M_M$) will be
different, since in one of them total fermion number is violated by 2
units.

When the Higgs scalars $\chi_L$ and $\chi_R$ acquire VEVs, the
first two operators ${\cal O}_1$ and ${\cal O}_2$ give the quark
masses:
\begin{eqnarray}
m_q ~\overline{q_L} ~q_R, \quad {\rm with} \quad m_q = {v_L v_R \over
M_D}\,. \end{eqnarray}
Similarly the third and the fourth operators ${\cal
O}_3$ and ${\cal O}_4$ contribute to the charged lepton and
neutrino Dirac masses:
\begin{eqnarray}
m_\ell ~\overline{\ell_L} ~\ell_R, \quad {\rm with} \quad m_\ell = {v_L v_R \over
M'_D}\,,  \end{eqnarray}
while the last two operators ${\cal O}_5$ and ${\cal
O}_6$ contribute to the Majorana masses for the left-handed and
right-handed neutrinos respectively:
\begin{eqnarray}
m_{\nu_L} = {v_L^2 \over M_M} \quad {\rm and} \quad m_{\nu_R} = {v_R^2 \over
M_M}\,. \end{eqnarray}
We shall now discuss some of the possible origin of these
operators and their consequences.

The see-saw masses of the neutrinos in theories with only doublet
Higgs may arise from various cases as, some higher dimensional
effective operators in a non supersymmetric theory, from
non-renormalizable gravitational interactions or from supersymmetric
extensions of models with doublet Higgs \cite{Brahmachari:2003wv}.  In
the present case the see-saw masses of the neutrinos can be obtained
in three different ways.  They may be mediated by exchange of scalar
fields or fermion fields or may be induced radiatively. As we shall
argue now, the first two possibilities are not very attractive and
hence we shall study the radiative mechanism in details.

When the intermediate field is a scalar, it has to be a field
which transforms as ${\bf 16 \times 16}$ and hence the field could
be either a ${\bf 10}$ or a ${\bf 120}$ or a ${\bf 126}$. If the
scalar field transform as ${\bf 120}$, the fermion mass matrix
will be totally antisymmetric and hence phenomenologically
unacceptable. If the scalar field $\Sigma$ transform as a ${\bf
10}$ or a ${\bf 126}$, its components will receive induced VEVs
through its couplings $\Sigma \Gamma \Gamma$ and $\Sigma
\Gamma^\dagger \Gamma^\dagger$. Then we can eliminate the $\chi_L$
and $\chi_R$ in the resulting theory and revert to the
conventional theories with bi-doublet Higgs $\phi$ and triplet
Higgs scalars $\Delta_{L,R}$. So, we shall not discuss this
possibility any further in the rest of the article.

We shall now consider the possibility of intermediate heavy fermions
generating the effective operators for the quark and lepton
masses. For each of the operators we require two fermions, one
left-handed and the other right-handed, both having same gauge
transformation properties. For the Majorana mass terms generated by
the last two operators a self-conjugate singlet fermion is
sufficient. The singlet fermion $S_{iL}$, we already included in the
present model, can give the Majorana masses to the left-handed and
right-handed neutrinos.

To generate the operator ${\cal O}_1$, we need two fermions $U_L$
and $U_R$ coupling to $ \overline{q_L} \chi_L$ and $q_R
\chi^*_R$ respectively. Both these fields should then transform
similarly $U_{L,R} \equiv ({\bf 3,1,1,}4/3) \subset ({\bf 15,1,1}) \subset
{\bf 45}$ or {\bf 210} and the Lagrangian must contain the couplings
\begin{equation}
{\cal L}_1 = a_1 \overline{U_L} q_R {\chi^*_{R}} + b_1
\overline{q_L} U_R {\chi_L} + m_{UU} \overline{U_L} U_R + h.c.\,,
\end{equation}
to give masses to the up-quarks by the operator ${\cal O}_1$. The
down quark masses are obtained by an effective operator ${\cal
O}_2$, which may be generated by adding the field $D_{L,R} \equiv
({\bf 3,1,1,}-2/3) \subset ({\bf 6,1,1}) \subset {\bf 10}$ or
{\bf 126} or
$D_{L,R} \equiv ({\bf 3,1,1,}-2/3) \subset ({\bf 10,1,1}) \subset {\bf 120}$
and
introducing the couplings in the Lagrangian:
\begin{equation}
{\cal L}_2 = a_2 \overline{D_L} q_R {\chi_R} + b_2 \overline{q_L}
D_R {\chi^*_L} + m_{DD} \overline{D_L} D_R + h.c.\,.
\end{equation}
The operators ${\cal O}_3$ and ${\cal O}_4$ may be obtained by
introducing the fields $N_{L,R} \equiv ({\bf 1,1,1,}0) \subset ({\bf 1,1,1})
\subset {\bf 1}$ or {\bf 45} and $E_{L,R} \equiv ({\bf 1,1,1,}-2) \subset
({\bf 10,1,1})
\subset {\bf 120}$ with the couplings
\begin{eqnarray}
{\cal L}_\ell &=& a_3 \overline{N_L} \ell_R {\chi^*_R} + b_3
\overline{\ell_L} N_R {\chi_L} + m_{NN} \overline{N_L} N_R
\nonumber \\
&&+\, a_4 \overline{E_L} \ell_R {\chi_R} + b_4 \overline{\ell_L} E_R
{\chi^*_L} + m_{EE} \overline{E_L} E_R + h.c.\,,
\end{eqnarray}
respectively. Then we may give masses to the up and the down
quarks as well as to the charged leptons and the neutrinos if
there are heave fermions transforming as {\bf 120} and {\bf 45}. The
singlet field $S_{iL}$ per generation is required to give Majorana
masses to the neutrinos with its couplings, which we shall discuss
later.

\begin{figure}[!h]
\begin{center}
\epsfxsize8cm\epsffile{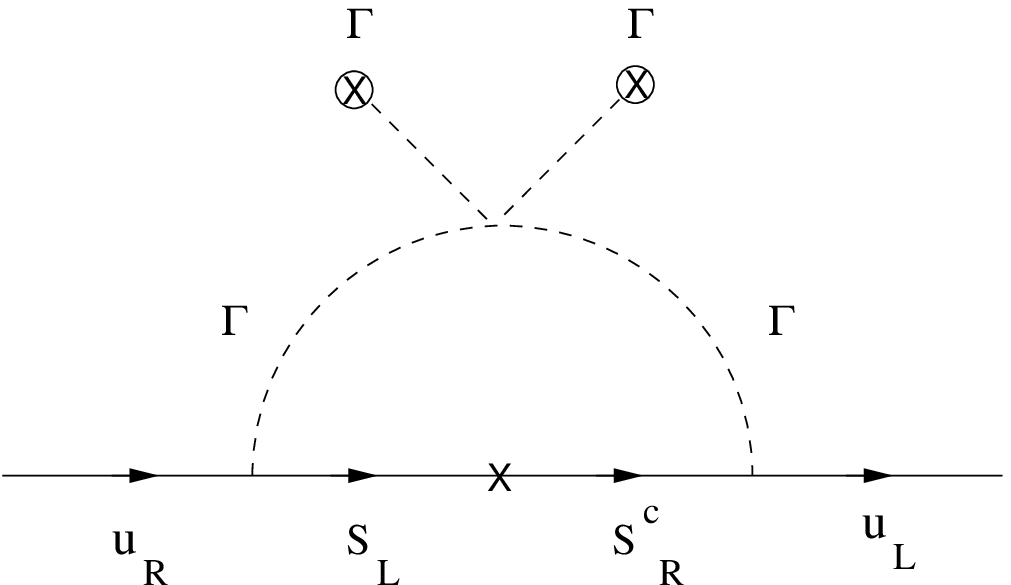} \caption{One loop diagram
contributing to the fermion masses.}
\label{fmass1.f}
\end{center}
\end{figure}

\begin{figure}[!ht]
\begin{center}
\epsfxsize8cm\epsffile{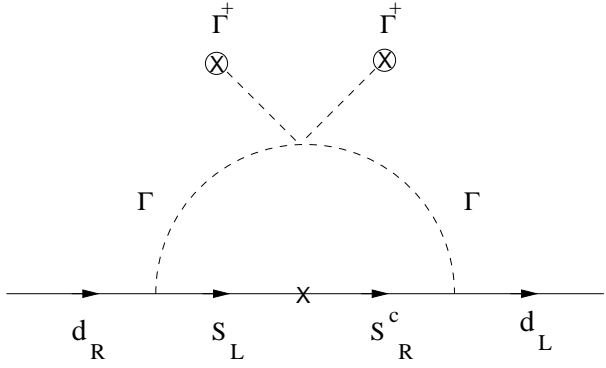} \caption{One loop diagram
contributing to the fermion masses.}
\label{fmass2.f}
\end{center}
\end{figure}

We shall now come back to the present model, where the quark and
lepton masses are generated radiatively. The fermion content of
the model has been discussed in sec. \ref{sbpat}. The most general Yukawa
couplings are then given by,
\begin{eqnarray}
{\cal L}_Y = f S_L \psi_L \Gamma^\dagger + M_S S_L S_L + h.c.\,.
\label{yukawac}
\end{eqnarray}
In this expression generation indices have been suppressed. One
loop diagram of Fig.~\ref{fmass1.f} then generates effective operators
\begin{equation}
\overline{\psi_L}~ \psi_R ~ \chi_L ~\chi^*_R \subset
\overline{\psi_L}~ \psi_R ~ \Gamma ~\Gamma\,, \label{xx}
\end{equation}
which are of the form ${\cal O}_1$ and ${\cal O}_3$ and
contributes to the down-quark and charged-lepton masses. On the
other hand the one loop diagram of Fig.~\ref{fmass2.f} generates effective
terms:
\begin{equation}
\overline{\psi_L}~ \psi_R ~ {\chi_R} ~{\chi^*_L} \subset
\overline{\psi_L}~ \psi_R ~ \Gamma^\dagger ~\Gamma^\dagger\,,
\end{equation}
which are of the form of the operators ${\cal O}_2$ and ${\cal
O}_4$ and contributes to the masses of the up-quarks and the Dirac
masses of the neutrinos.

The up-quark, down-quark and charged-lepton masses can now be
estimated from Fig.~\ref{fmass1.f} and Fig.~\ref{fmass2.f} to be:
\begin{eqnarray}
M_u &=& {\lambda'_\Gamma \over 8 \pi^2} { m_R m_L \over M_X}\,,
\label{mup} \\
M_{d,\ell} &=& {\lambda_\Gamma \over 8 \pi^2} { m_R m_L
\over M_X}\,.
\label{mdwn}
\end{eqnarray}
Here $M_X = M_\Gamma^2 /M_S$ or $M_S$, depending on whether
$M_\Gamma$ or $M_S$ is larger and $ m_L = f v_L$, and
$m_R = f v_R .$ We thus obtain different up and down quark
masses and on the other hand $b - \tau$ unification. The other
mass relations in the down-quark sector and the charged-lepton
mass relations could come from higher order terms, since the
remaining matrix elements are of the order of $10^{-3}$ to
$10^{-5}$ compared to the 33-element \cite{ss1,ss2}. For example,
operators of the form
\begin{equation}
\overline{\psi_L}~ \psi_R ~ \Gamma ~\Gamma ~\Phi_4 \quad {\rm and}
\quad \overline{\psi_L}~ \psi_R ~ \Gamma^\dagger ~\Gamma^\dagger
~\Phi_4\,, 
\end{equation}
contribute differently to the down-quark and charged-lepton
masses, since the effective VEV transform as $\Gamma \Gamma
\Phi_4$ and $\Gamma^\dagger ~\Gamma^\dagger \Phi_4$, which behaves
as the field ${\bf (15,2,2) \subset 126}$ and hence can solve the
fermion mass problem in GUTs, a la Georgi-Jarlskog mechanism.

The neutrino masses come from the couplings of the neutrinos with
the singlet fermions $S_{iL}$, given by Eq.~(\ref{yukawac}).
In the basis $\pmatrix{ \nu_L & {\nu^c}_L & S_L }$ the tree level
neutrino mass matrix becomes:
\begin{equation}
M_\nu = \pmatrix{ 0 & 0 & m_L \cr 0 & 0 &m_R  \cr m_L & m_R & M_S
} ,
\end{equation}
which gives two heavy states, which are mostly $S_L$ and
${\nu^c}_L$. In the limit $M_S \gg m_R$, the two heavy mass
eigenvalues are $M_S$ and $m_R^2/M_S$. On the other hand, when
$M_S \ll m_R$, the two heavy states are almost degenerate with
eigenvalues $\pm m_R$ with a mass splitting of about $M_S$. The
latter case may be more interesting for leptogenesis, which we shall
discuss at the end.

\begin{figure}[!ht]
\begin{center}
\epsfxsize12cm\epsffile{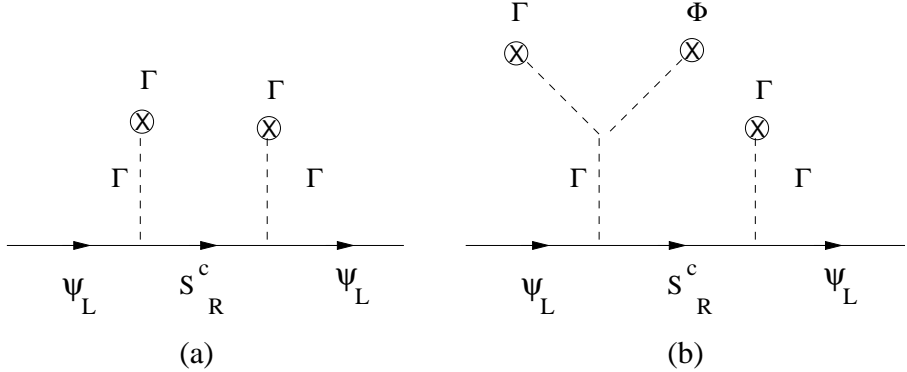}
\caption{Tree level diagrams
contributing to the neutrino masses.}
\label{xyfig3}
\end{center}
\end{figure}

The lightest state $\nu_L$ remains massless at the tree level.
However, if we include the effect of $D$-parity violation, this
problem could be solved. We thus continue our discussion taking
$D$-parity violation into consideration. The effective operator:
\begin{equation}
\nu_L~ \nu_L~ {\chi^*_L}~ {\chi^*_L} + {\nu^c}_L~ {\nu^c}_L~
{\chi_R} ~{\chi_R} + {\nu^c}_L~ \nu_L ~{\chi_R}
~{\chi^*_L} \subset \psi_L~ \psi_L ~ \Gamma^\dagger
~\Gamma^\dagger  + h.c.\,,
\label{xx1}
\end{equation}
and a similar $D$-parity violating effective operator
\begin{equation}
\alpha\,\psi_L~ \psi_L ~ \Gamma^\dagger ~\Gamma^\dagger ~\Phi +
h.c. \supset - \alpha\, \nu_L~ \nu_L~ {\chi^*_L}~ {\chi^*_L} ~\Phi_D +
\alpha {\nu^c}_L~ {\nu^c}_L~ {\chi_R} ~{\chi_R} ~\Phi_D\,,
\label{xx2}
\end{equation}
which could come from the Fig.~\ref{xyfig3}a and Fig.~\ref{xyfig3}b,
together give a neutrino mass matrix:
\begin{equation}
M_\nu = \pmatrix{ (1-\alpha) {m_L^2 \over M_S} & {m_L m_R \over
M_S} \cr {m_L m_R \over M_S} & (1+\alpha) {m_R^2 \over M_S}} ,
\end{equation}
where, $ \alpha = M_D ~{\langle \Phi_D \rangle / m_\Phi^2}$. This
mass matrix is obtained by integrating out the heavy modes $S_i$.
In the absence of $D$-parity violation, this mass matrix remains
symmetrical and one of the eigenvalues vanishes, leading to a
massless left-handed neutrino. When $D$-parity violating effect is
included, the symmetry between the left and the right handed
neutrinos is lost and the left-handed neutrinos become light and
massless. In the limit $\alpha >-1$ and $m_L \ll m_R$,
diagonalization of this matrix gives a light neutrino with mass
\begin{equation}
m_\nu =  { \alpha^2 \over (1 + \alpha)}  {m_L^2 \over M_S}\, .
\end{equation}
This gives the correct order of magnitude for neutrino mass for $m_L \sim 100$ GeV and
$M_S \sim 10^{13}$ GeV. This tiny neutrino mass is of the see-saw
type and in fact all fermion masses are of the see-saw type in
this model.
\section{Gauge coupling unification}
\label{gcu}
In this section we shall study the renormalization group equations
for the evolution of the coupling constants in our model. We start
with the one-loop renormalization group equation for the gauge
coupling constants
\begin{equation}
\frac{d g_i}{d t } = \beta_i\,,
\end{equation}
where $t=\ln(\mu)$ where $\mu$ stands for the energy-scale of our
theory. $g_i$ is the gauge coupling constant of the group $G_i$ which
is a subgroup of the semi-simple gauge group $G_1\times G_2 \times
\cdot \cdot \times G_i \times \cdot \cdot \cdot$ and the beta
functions $\beta_i$ contain contributions from gauge bosons, fermions
and scalars as:
\begin{eqnarray}
  \beta_i &=& {g_i^3 \over 16 \pi^2} \left[ \{\rm gauge ~bosons \} + \{\rm fermions\} + \{\rm scalars \} \right].
\end{eqnarray}
To two-loop the $\beta$ functions of any semi-simple gauge group is
given as \cite{Jones:1981we}:
\begin{eqnarray}
\beta_i=\frac{a_i g^3_i}{16\pi^2} +
\frac{1}{(16\pi^2)^2}\sum^{n}_{j=1}b_{ij}g^2_j g^3_i\,,
\label{gbeta}
\end{eqnarray}
where the $a_i$s and $b_{ij}$s are the one-loop and two-loop
$\beta$ function coefficients respectively.
Here $n$ is the number of groups whose direct product is the
semi-simple gauge group of the theory, $i,j$ takes on values from
$1,2,\cdot \cdot\cdot n$.

First we concentrate on the one-loop effect and later we will see the
effects of the two-loop coefficients on the gauge coupling evolutions.
The $a_i$s calculated for the various phases are supplied below 
\cite{Jones:1981we}.
\begin{eqnarray}
\begin{tabular}{|c|c|c|}
\hline
${\rm G}_{\rm SM}$ & ${\rm G}_{3221}$ & ${\rm G}_{422}$ \\
\hline
  & & \\
$a^{(SM)}_{1Y}=\frac{23}{5}$ & $a^{(LR)}_{B-L}=5$ & $a^{(X)}_{2L}=-2$\\
  & & \\
$a^{(SM)}_{2L}=-3$ & $a^{(LR)}_{2L}=-3$ & $a^{(X)}_{2R}=-2$\\
  & & \\
$a^{(SM)}_{3C}=-7$ & $a^{(LR)}_{2R}=-3$ & $a^{(X)}_{4C}=-8$\\
  & & \\
        & $a^{(LR)}_{3C}=-7$ &            \\
  & & \\
\hline
\end{tabular}
\nonumber
\end{eqnarray}
In the above table ${\rm G}_{\rm SM}={\rm SU(3)}_c\times{\rm
SU(2)}_L\times {\rm U(1)}_{\rm em}$ and the superscripts $(SM)$,
$(LR)$, $(X)$ indicates the phase in which the numbers are
calculated. The renormalization group (RG) equations can now be
used to write down the the standard model gauge couplings in terms
of the SO(10) coupling. Writing $\alpha_i = g^2_i/4\pi$,
the gauge coupling constant matching conditions at the $M_R$ scale
are:
\begin{eqnarray}
\left[\frac{1}{\alpha_{1Y}(M_R)}\right]_{{\rm G}_{\rm SM}}&=&
\left[\frac35\frac{1}{\alpha_{2R}(M_R)} + \frac23
\frac{1}{\alpha_{B-L}(M_R)}\right]_{{\rm G}_{3221}} ,\\
\left[\frac{1}{\alpha_{2L}(M_R)}\right]_{{\rm G}_{\rm SM}}&=&
\left[\frac{1}{\alpha_{2L}(M_R)}\right]_{{\rm G}_{3221}} ,\\
\left[\frac{1}{\alpha_{2L}(M_R)}\right]_{{\rm G}_{\rm SM}}&=&
\left[\frac{1}{\alpha_{2R}(M_R)}\right]_{{\rm G}_{3221}} ,\\
\left[\frac{1}{\alpha_{3C}(M_R)}\right]_{{\rm G}_{\rm SM}}&=&
\left[\frac{1}{\alpha_{3C}(M_R)}\right]_{{\rm G}_{3221}} .
\label{mmr}
\end{eqnarray}
The matching conditions at the $M_X$ scale are:
\begin{eqnarray}
\left[\frac{1}{\alpha_{B-L}(M_R)}\right]_{{\rm G}_{3221}}&=&
\left[\frac{1}{\alpha_{4C}(M_R)}\right]_{{\rm G}_{422}} ,\\
\left[\frac{1}{\alpha_{2L}(M_R)}\right]_{{\rm G}_{3221}}&=&
\left[\frac{1}{\alpha_{2L}(M_R)}\right]_{{\rm G}_{422}} ,\\
\left[\frac{1}{\alpha_{2R}(M_R)}\right]_{{\rm G}_{3221}}&=&
\left[\frac{1}{\alpha_{2R}(M_R)}\right]_{{\rm G}_{422}} ,\\
\left[\frac{1}{\alpha_{3C}(M_R)}\right]_{{\rm G}_{3221}}&=&
\left[\frac{1}{\alpha_{4C}(M_R)}\right]_{{\rm G}_{422}} .
\label{mmx}
\end{eqnarray}
Finally at the $M_U$ scale,
\begin{eqnarray}
\left[\frac{1}{\alpha_{4C}(M_U)}\right]_{{\rm G}_{422}}&=&
\left[\frac{1}{\alpha_{10}(M_U)}\right]_{{\rm SO}_{10}} ,\\
\left[\frac{1}{\alpha_{2L}(M_U)}\right]_{{\rm G}_{422}}&=&
\left[\frac{1}{\alpha_{10}(M_U)}\right]_{{\rm SO}_{10}} ,\\
\left[\frac{1}{\alpha_{2R}(M_U)}\right]_{{\rm G}_{422}}&=&
\left[\frac{1}{\alpha_{10}(M_U)}\right]_{{\rm SO}_{10}} .
\label{mmu}
\end{eqnarray}
With the help of the above matching conditions and the RG equation we can
write to one-loop,
\begin{eqnarray}
\frac{1}{\alpha_{1Y}(M_Z)}&=&\frac{1}{\alpha_{10}(M_U)} + 8\pi\left[
a^{(SM)}_{1Y}\ln\left(\frac{M_R}{M_Z}\right) + \left(\frac35
a^{(LR)}_{2R} + \frac{2}{5}
a^{(LR)}_{B-L}\right)\ln\left(\frac{M_X}{M_R}\right)\right.\nonumber\\
&&+\left.\left(\frac35 a^{(X)}_{2R} + \frac{2}{5}
a^{(X)}_{B-L}\right)\ln\left(\frac{M_U}{M_X}\right)\right]\,,
\label{aly}\\
\frac{1}{\alpha_{2L}(M_Z)}&=&\frac{1}{\alpha_{10}(M_U)} + 8\pi\left[
a^{(SM)}_{2L}\ln\left(\frac{M_R}{M_Z}\right) + a^{(LR)}_{2L}
\ln\left(\frac{M_X}{M_R}\right)\right.\nonumber\\
&&+\left.a^{(X)}_{2L}\ln\left(\frac{M_U}{M_X}\right)\right],
\label{al2l}\\
\frac{1}{\alpha_{3C}(M_Z)}&=&\frac{1}{\alpha_{10}(M_U)} + 8\pi\left[
a^{(SM)}_{3C}\ln\left(\frac{M_R}{M_Z}\right) + a^{(LR)}_{3C}
\ln\left(\frac{M_X}{M_R}\right)\right.\nonumber\\
&&+\left.a^{(X)}_{4C}\ln\left(\frac{M_U}{M_X}\right)\right].
\label{al3}
\end{eqnarray}
The linear combinations of the gauge couplings that yields $\sin^2
\theta_W$ and $\alpha_s$ are the following:
\begin{eqnarray}
\sin^2 \theta_W(M_Z) &=& \frac38 - \frac58 \alpha(M_Z)\left(\frac{1}{\alpha_1Y(M_Z)} -
\frac{1}{\alpha_{2L}(M_Z)}\right)\,,
\label{wang}\\
1-\frac83\frac{\alpha(M_Z)}{\alpha_s(M_Z)}&=&\alpha(M_Z)\left(\frac{1}{\alpha_{2L}(M_Z)} +
\frac{5}{3\alpha_1Y(M_Z)} - \frac{8}{3\alpha_3C(M_Z)}\right)\,,
\label{as}
\end{eqnarray}
where $\alpha$ and $\alpha_s$ are related to the electromagnetic and
strong interaction coupling constants in the present symmetry broken phase.
Using the experimental numbers \cite{amal,anse},
\begin{eqnarray}
\sin^2 \theta_W(M_Z) = 0.2312,\quad\alpha^{-1}(M_Z)=128.91 \quad{\rm and}\quad\alpha_s(M_Z)=0.1187\,.
\end{eqnarray}
Eq.~(\ref{wang}) and Eq.~(\ref{as}) reduces to the following:
\begin{eqnarray}
\frac{1}{\alpha_{1Y}(M_Z)} -
\frac{1}{\alpha_{2L}(M_Z)} &=& 29.66\,,
\label{cons1}\\
\frac{1}{\alpha_{2L}(M_Z)} +
\frac{5}{3\alpha_{1Y}(M_Z)} - \frac{8}{3\alpha_{3C}(M_Z)} &=& 106.444\,.
\label{cons2}
\end{eqnarray}
The above equations can be utilized for calculating the
intermediate scales like $M_R$ and $M_X$ in our theory. Here we
discuss two cases.
\subsubsection*{When $M_X=M_U$}
In this case from Eq.~(\ref{aly}), Eq.~(\ref{al2l}) and Eq.~(\ref{cons1})
and using the $\beta$ function coefficients given in the last table we get,
\begin{eqnarray}
\frac{1}{\alpha_{1Y}(M_Z)} -
\frac{1}{\alpha_{2L}(M_Z)}=\frac{1}{2\pi}\left[\frac{23}{5}\ln\left(\frac{M_R}
{M_Z}\right) + \frac15\ln\left(\frac{M_U}{M_R}\right) + 3\ln\left(\frac{M_U}
{M_Z}\right)\right]=29.66\,.\nonumber\\
\label{mumx1}
\end{eqnarray}
Similarly from Eq.~(\ref{aly}), Eq.~(\ref{al2l}), Eq.~(\ref{al3})
and Eq.~(\ref{cons2}) and the $\beta$ function coefficients we
get,
\begin{eqnarray}
\frac{1}{\alpha_{2L}(M_Z)} +
\frac{5}{3\alpha_{1Y}(M_Z)} - \frac{8}{3\alpha_{3C}(M_Z)}&=&
\frac{1}{2\pi}\left[15\ln\left(\frac{M_U}{M_Z}\right) +
\frac{23}{3}\ln\left(\frac{M_R}{M_Z}\right)\right.\nonumber\\
&&+\left. \frac13\ln\left(\frac{M_U}{M_R}\right)
\right]=106.444\,.
\label{mumx2}
\end{eqnarray}
Eliminating $M_U$ from the above two equations we get
\begin{eqnarray}
\ln\left(\frac{M_R}{M_Z}\right)=16.301\,,
\label{mremw}
\end{eqnarray}
and if we take $M_Z= 91.1876$ GeV then
$M_R=1.1\times 10^9$ GeV.
The above value of $M_R$ can be taken as the lowest possible value of it 
in our model and all the predictions in our model will be made assuming 
$M_R \gg 10^9$ GeV.    
\subsubsection*{When $M_X\neq M_U$}
In this case the two equations corresponding to Eq.~(\ref{mumx1}) and
Eq.~(\ref{mumx2}) are:
\begin{eqnarray}
\frac{1}{2\pi}\left[\frac{23}{5}\ln\left(\frac{M_R}
{M_Z}\right) + \frac15\ln\left(\frac{M_X}{M_R}\right) -
\frac{12}{5}\ln\left(\frac{M_U}
{M_X}\right)+ 3\ln\left(\frac{M_X}{M_Z}\right)\right]=29.66\,,
\label{munmx1}
\end{eqnarray}
and
\begin{eqnarray}
\frac{1}{2\pi}\left[\frac{23}{3}\ln\left(\frac{M_R}
{M_Z}\right) + \frac13\ln\left(\frac{M_X}{M_R}\right) -
12\ln\left(\frac{M_U}
{M_X}\right)+ 15\ln\left(\frac{M_X}{M_Z}\right)\right]=106.444\,.
\label{munmx2}
\end{eqnarray}
Eliminating $M_U$ from the above two equations gives us a relation between
$M_X$ and $M_R$ as,
\begin{eqnarray}
\ln\left(\frac{M_X}{M_R}\right)=21\left[\ln\left(\frac{M_R}{M_Z}\right)
-19.214\right]\,.
\label{mrmxp}
\end{eqnarray}
From the above equation it can be verified that if we take
$M_Z= 91.1876$ GeV and impose $M_R=M_X$ then $M_R= 2.01\times 10^{10}$
GeV. In the next subsection we include the two loop results and the
above results are re-derived computationally. From the computational
results we see that the above value of $M_R\sim 10^{11}$ GeV is
two orders of magnitude smaller than the actual one.  
%
\subsection*{Two-loop result}
After the discussion on gauge coupling unification to one-loop we
discuss about the two-loop effects of the RG equations. To two-loop
the $\beta$ functions are given in Eq.~(\ref{gbeta}).  The $a_i$s for
the various phases has been supplied in the table appearing in the
beginning of this section and the $b_{i j}$s for the various phases
are as follows:
\begin{eqnarray}
b^{(SM)}_{ij}=
\left(
\begin{array}{ccccc}
\frac{176}{25}& &\frac{18}{5}& &\frac{44}{5}\\
 & & & &\\
\frac{6}{5}& & 8 & &12\\
 & & & &\\
\frac{11}{10}& &\frac{9}{2}& &-26
\end{array}
\right)\,.
\label{bijsm}
\end{eqnarray}
Here $i,j=1Y,\,\,2L,\,\,3C$,
\begin{eqnarray}
b^{(LR)}_{ij}=
\left(
\begin{array}{ccccccc}
8& &9& &9& &2\\
3& &8& &0& &12\\
3& &0& &8& &12\\
\frac12& &\frac92& &\frac92& &-26
\end{array}
\right)\,,
\label{bijlr}
\end{eqnarray}
and here $i,j=B-L,\,\,2L,\,\,2R,\,\,3C$,
\begin{eqnarray}
b^{(X)}_{ij}=
\left(
\begin{array}{ccccc}
\frac{199}{3}& &0& &\frac{105}{2}\\
 & & & & \\
0& &\frac{199}{3}& &\frac{105}{2}\\
 & & & & \\
\frac{21}{2}& &\frac{21}{2}& &\frac{1307}{6}
\end{array}
\right)\,.
\label{bijX}
\end{eqnarray}
Where $i,j=2L,\,\,2R,\,\,4C$.
%
%
\begin{figure}[!h]
\centering
\includegraphics[height=165mm,keepaspectratio=true]{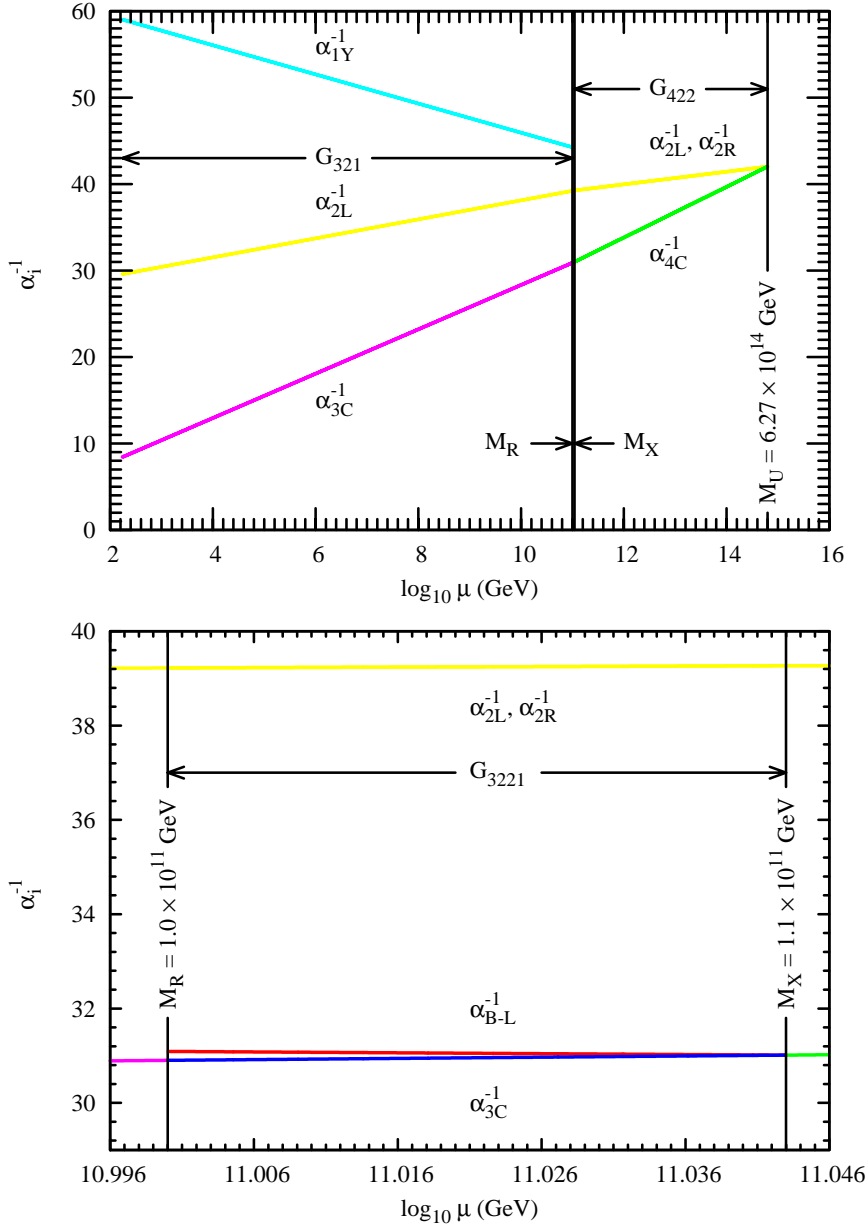}
\caption{\small\sf Plot showing the evolution of $\alpha_i^{-1}$ s 
in the SM phase, left-right phase upto $M_U$. The abscissa is $\ln(\mu)$, where $\mu$ initial is
$M_t = 173$ GeV and $M_R=10^{11}$ GeV in our case. The figure in the right shows
explicitly the gauge coupling unification at $M_U\sim 10^{15}$ GeV.}
\label{gcsm:f}
\end{figure}

If we start from $\mu=M_T=173$ GeV (where $n_f=6$) and fix $M_R=10^{11}$ GeV then
the evolution of the gauge coupling constants are as given in
Fig.~\ref{gcsm:f}.  In the next phases the coupling constant evolution
shows that at $M_X$ both $\alpha_{3C}$ and $\alpha_{B-L}$ unite to produce
$\alpha_{4C}$. Fig.~\ref{gcsm:f} shows that from $M_X$ on wards the
development of $\alpha_{2L}$ and $\alpha_{2R}$ are identical.  At
around $M_U\sim 10^{15}$ GeV the gauge coupling constants unite.
Our computational results show that the highest value of $M_R$ is just
slightly above $10^{11}$ GeV. If $M_R$ is much above the above mentioned
value then $M_X$ comes down and $M_R \ge M_X$.

\begin{figure}[!h]
\centering
\includegraphics[height=140mm,keepaspectratio=true,angle=-90]{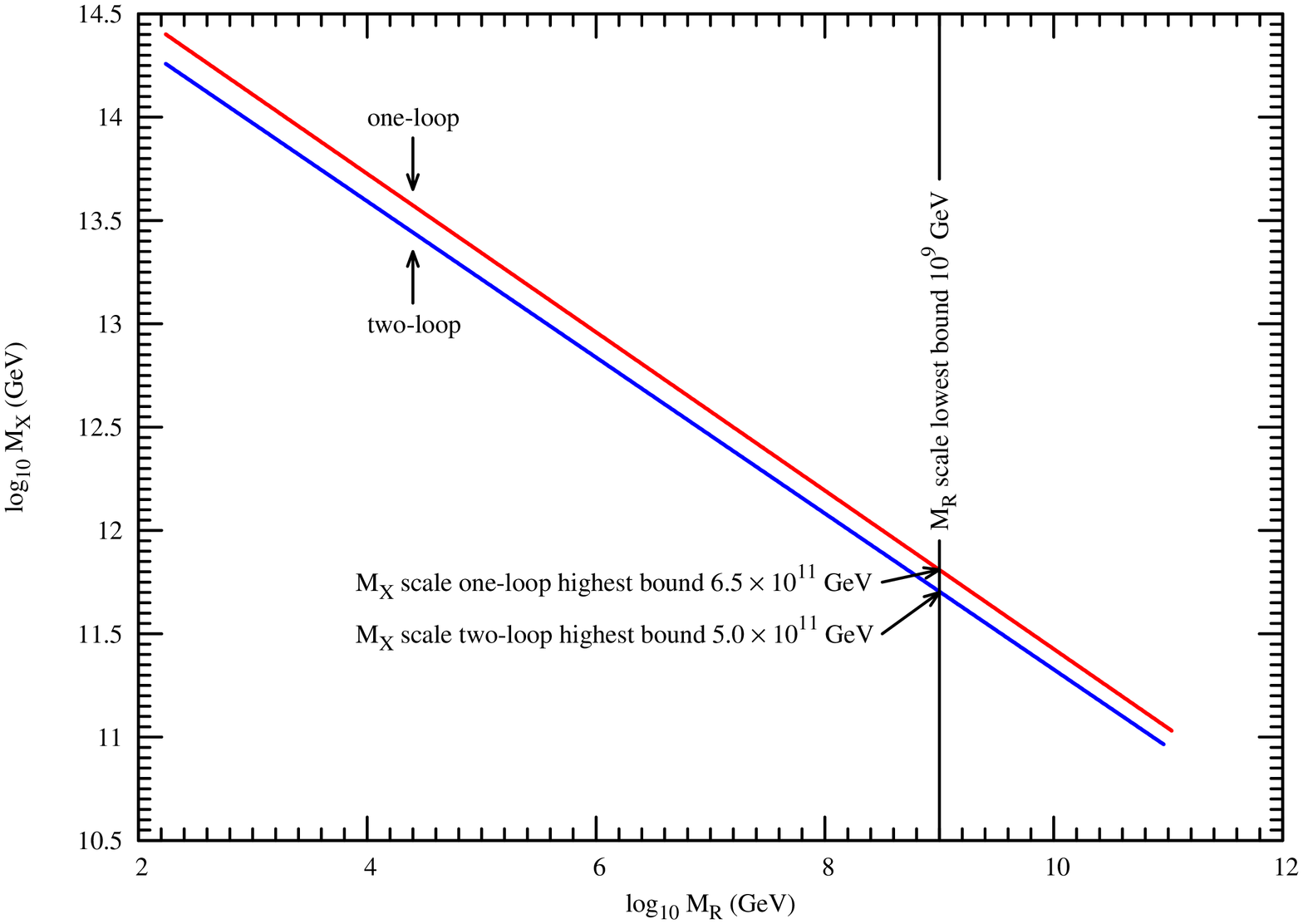}
\caption{$M_R$ vs. $M_X$ shows lowest and highest bound.}
\label{bound1}
\end{figure}

\begin{figure}[h]
\centering
\includegraphics[height=140mm,keepaspectratio=true,angle=-90]{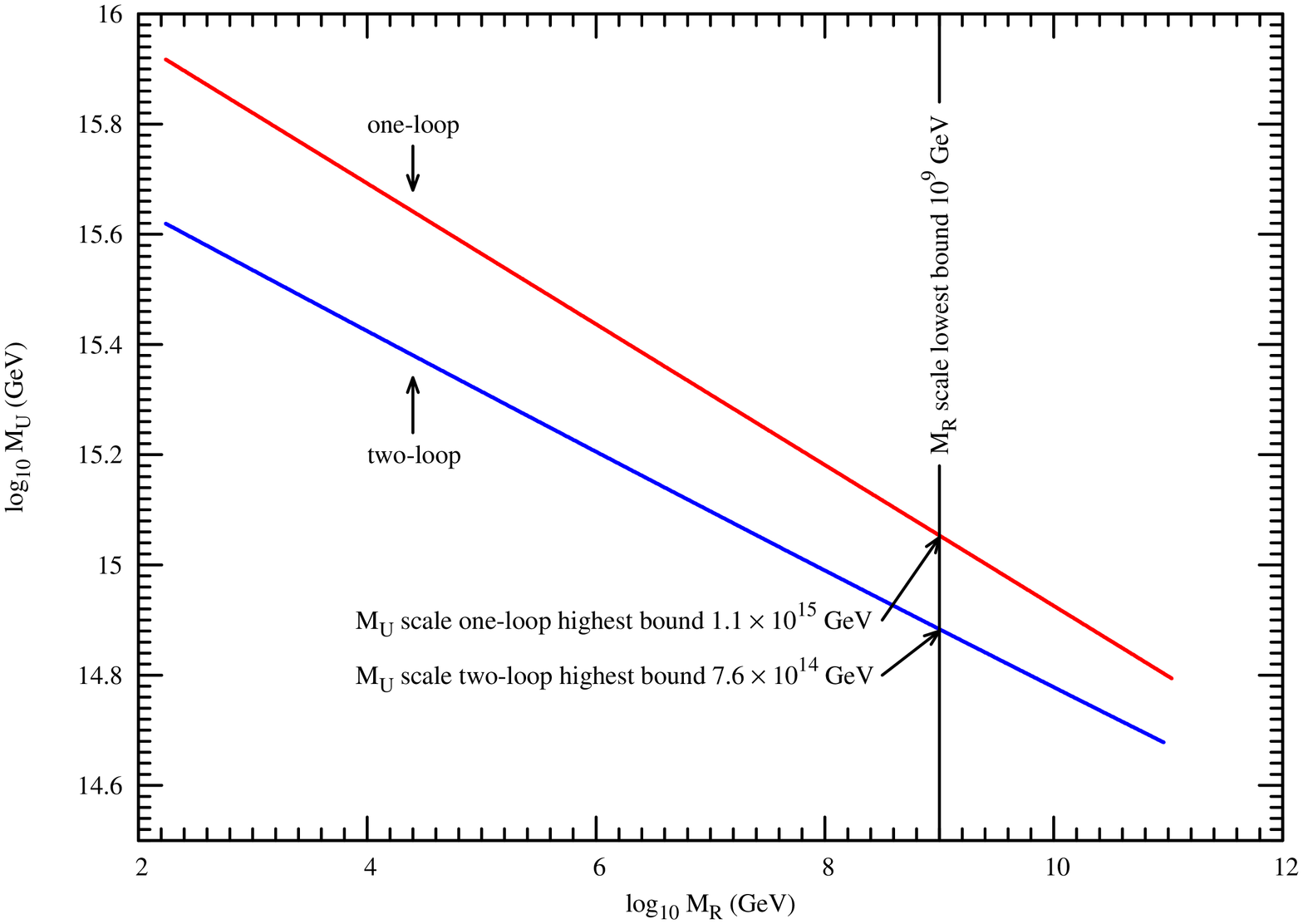}
\caption{$M_R$ vs. $M_U$ shows lowest and highest bound.}
\label{bound2}
\end{figure}

\begin{figure}[h]
\centering
\includegraphics[height=140mm,keepaspectratio=true,angle=-90]{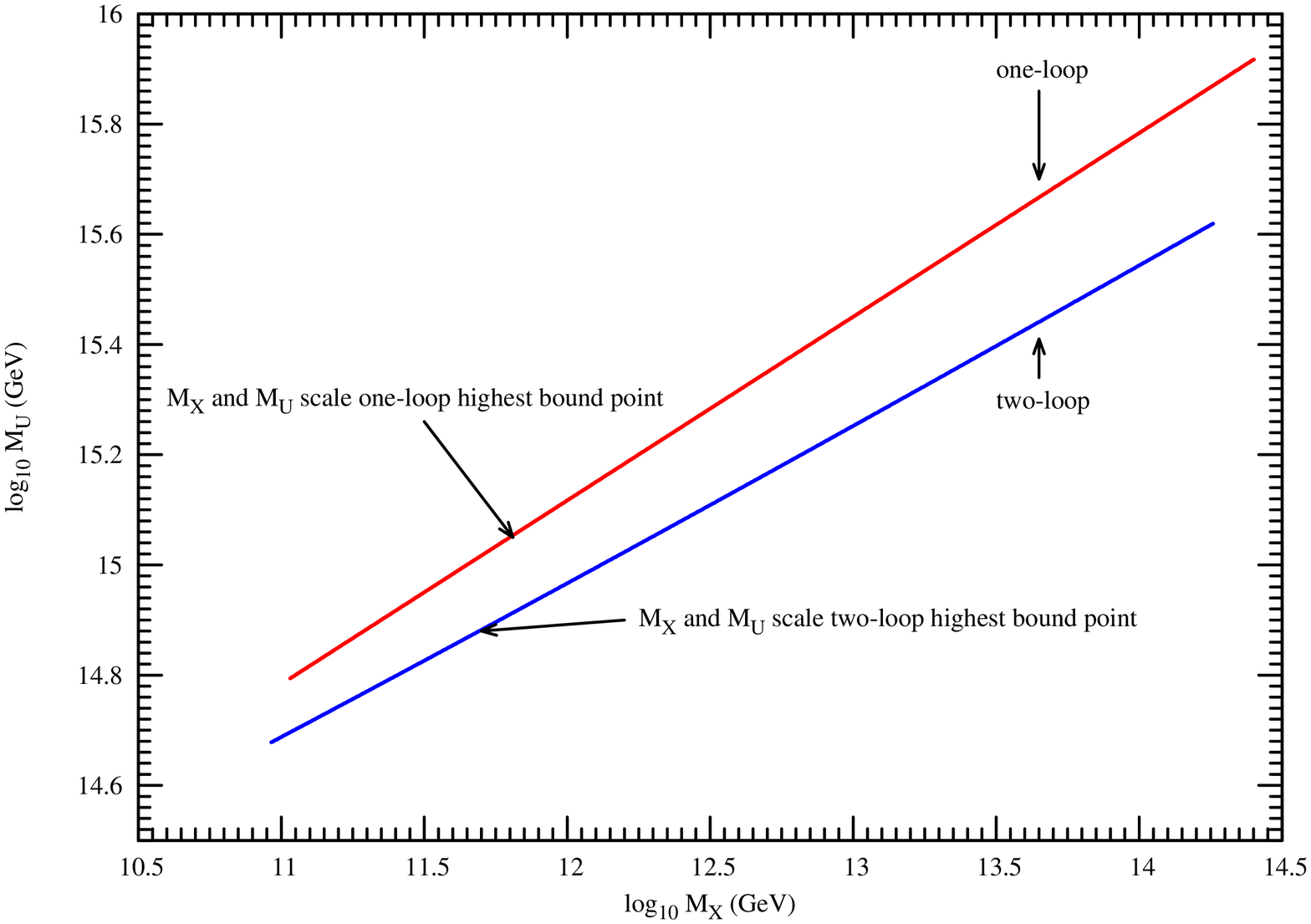}
\caption{$M_X$ vs. $M_U$ shows lowest and highest bound.}
\label{bound3}
\end{figure}

\begin{figure}[h]
\centering
\includegraphics[height=140mm,keepaspectratio=true,angle=-90]{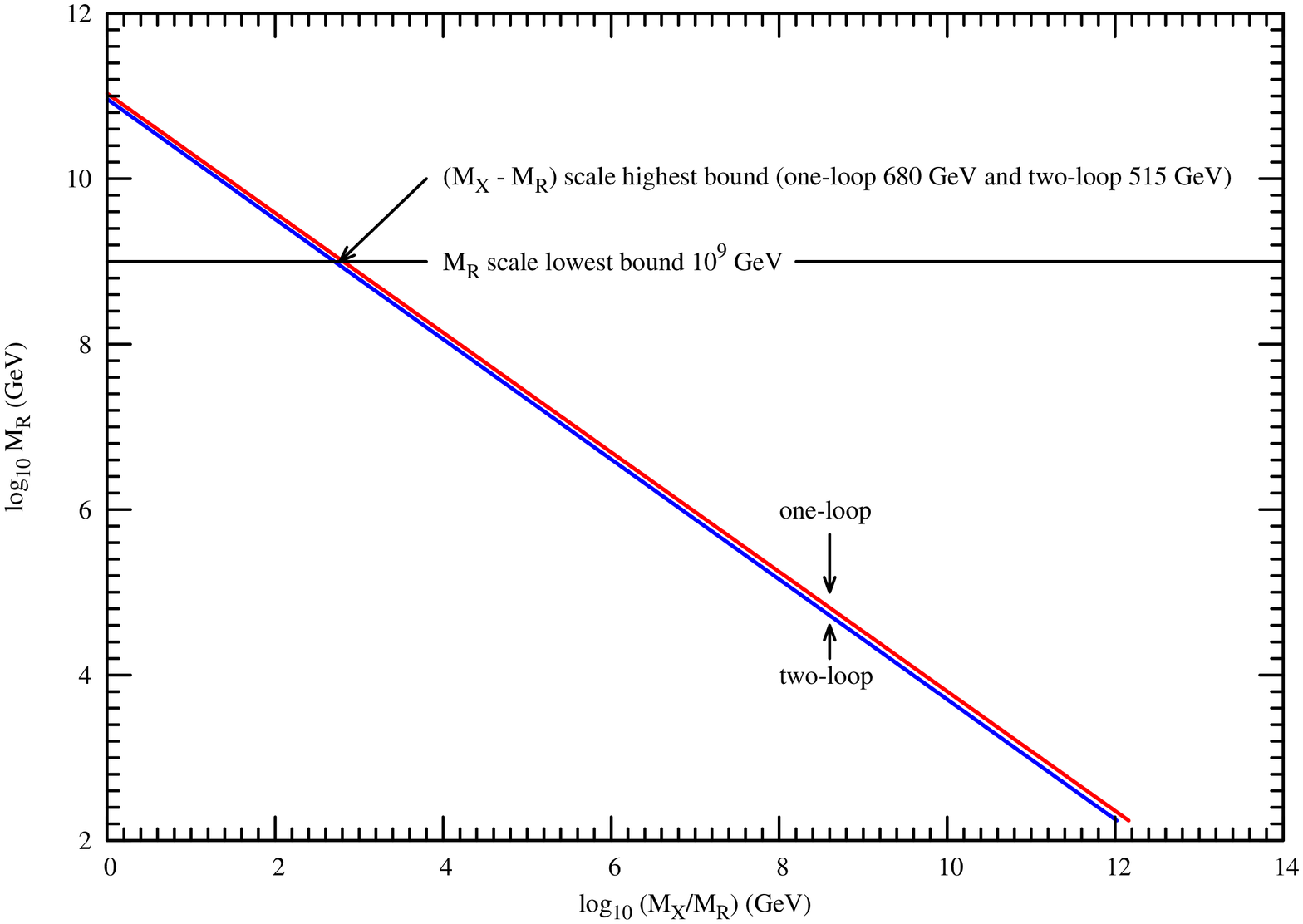}
\caption{$M_X-M_R$ vs. $M_X$ shows how $M_X-M_R$ approaching zero.}
\label{bound4}
\end{figure}

In models with triplet Higgs scalars, it is possible to consider $M_R
\geq M_X$.  At the scale $M_R$, the group ${\rm SU(2)}_R$ is broken
into ${\rm U(1)}_R$ and stays orthogonal to ${\rm SU(4)}_C$. Later
when ${\rm SU(4)}_C$ is broken to ${\rm SU(3)}_C \times {\rm
U(1)}_{B-L}$. Subsequently at a much lower scale the symmetry
breaking, ${\rm U(1)}_R \times {\rm U(1)}_{B-L} \to {\rm U(1)}_Y$,
takes place.  However, this is not possible in scenarios with only
doublet Higgs scalars, since the right-handed Higgs scalar doublet
does not have any component with $B-L$ quantum number to be zero and
hence it breaks ${\rm SU(2)}_R$ and ${\rm U}(1)_{B-L}$
simultaneously. As a result, the highest value of $M_R$ could be
$10^{11}$ GeV. As we discussed earlier \cite{sridhar}, the gauge
coupling unification also requires a lower bound on $M_R > 10^9$
GeV. From this lower (higher) bound for $M_R$, we estimated
computationally higher (lower) bound for $M_X$ and $M_U$ as shown in
Figs.~\ref{bound1}, \ref{bound2}, \ref{bound3}. The lower and higher
bounds for $M_X$ are $\sim 10^{11}$ GeV and $\sim 5\times 10^{11}$ GeV
respectively.  Also lower and higher bounds for $M_U$ are estimated as
$5\times 10^{14}$ GeV and $\sim 10^{15}$ GeV respectively. In this
regard it is important to note that the stability of the proton offers
further constraints on the parameter space of their model, since the
GUT scale is lower than most of the conventional models and is smaller
than $10^(15)$GeV.

The difference between $M_X$ and $M_R$ can go up from 0 (i.e,
$M_R=M_X$; no intermediate $G_{3221}$ symmetry) to $\sim 550$ GeV as
shown in Fig.~\ref{bound4}.  This range of $M_R$ plays an important
role for Yukawa coupling unification and leptogenesis in this model.
\section{Yukawa coupling evolution}\label{yukawa}
In sec. \ref{fmass}, Eq.~(\ref{yukawac}) gives the only Yukawa
couplings of our theory. Eq.~(\ref{yukawac}) is valid in the SO(10) level.
In the ${\rm G}_{422}$ level the Yukawa couplings will become:
\begin{eqnarray}
{\cal L}^{(4,2,2)}_Y &=& g' S_L F^{(l)}_L \Phi^*_L + g'' S_L F^{(r)}_L \Phi_R
+M_S S_L S_L + h.c.\,.
\end{eqnarray}
The gauge transformation properties of the various fields, except $S_L$ which 
is a gauge singlet, are shown below:
\begin{eqnarray}
\begin{array}{ccccc}
F^{(l)}_L &\equiv& ({\bf 4,2,1})\subset{\bf 16} &\equiv& \Gamma\,,\\
F^{(r)}_L &\equiv& ({\bf \overline{4},1,2})\subset{\bf 16} &\equiv& \Gamma\,,\\
\Phi^*_L &\equiv& ({\bf \overline{4},1,2})\subset{\overline{\bf 16}} &\equiv& {\Gamma^\dagger}\,,\\
\Phi_{R} &\equiv& ({\bf 4,1,2})\subset{\overline {\bf 16}}
&\equiv& {\Gamma^\dagger}\,.
\end{array}
\end{eqnarray}
In the above equations $(\bf{x , y, z})$ designates the transformation 
properties under ${\rm G}_{422}$.

The Yukawa couplings in the ${\rm G}_{3221}$ phase is as:
\begin{eqnarray}
{\cal L}^{(3,2,2,1)}_Y &=& k'_{(q)} S_L q_L H^*_L + k'_{(l)} S_L
\ell_L \chi^*_L
+ k''_{(q)} S_L (q_R)^c H_R + k''_{(l)} S_L (\ell_R)^c \chi_R \nonumber\\
&&+\, M_S S_L S_L + h.c.\,.
\label{3221}
\end{eqnarray}
The gauge transformation properties of $q_L$, $\ell_L$, $(q_R)^c$, 
$(\ell_R)^c$, $\chi^*_L$ and $\chi_R$ are specified in Eq.~(\ref{lql}), 
Eq.~(\ref{rql}) and Eq.~(\ref{higgsconj}). Here we specify the gauge transformation properties of the other two fields present in the above equation.
\begin{eqnarray}
\begin{array}{cccccc}
H^*_L &\equiv& ({\bf \overline{3},2,1},-\frac{1}{3})
\subset({\bf \overline{4},2,1})\subset{\bf \overline{16}} 
&\equiv& \Gamma^\dagger\,,\\
H_{R} &\equiv& ({\bf 3,1,2},\frac{1}{3})\subset(\bf {4,1,2})
\subset {\bf 16} &\equiv& \Gamma\,.\\
\end{array}
\end{eqnarray}
In the above equations $({\bf x, y, z}, w)$ designates the ${\rm
G}_{3221}$ transformation properties and $(\bf{x, y, z})$ designates
the ${\rm G}_{422}$ transformation properties.

In the next stage, that is in the ${\rm G}_{321}$ phase, the Yukawa 
couplings look like:
\begin{eqnarray}
{\cal L}^{(3,2,1)}_Y &=& y_{(qL)} S_L Q_L T^*_L +
y_{(lL)} S_L L_L h^*_L + y_{(u)} S_L (Q_R)^c_u\, T^{(-)}_R\nonumber\\
&&+\, y_{(d)} S_L (Q_R)^c_d\, T^{(+)}_R + y_{(e)} S_L(e_R)^c\,
\phi^{(+)}_{*R}\nonumber\\
&&+\, y_{(\nu)} S_L(\nu_R)^c\,\phi^{(0)}_{*R}+ M_S S_L S_L + h.c.\,.
\label{321}
\end{eqnarray}
Here the various standard model fermion fields transform under 
${\rm G}_{321}$ as:
\begin{eqnarray}
\begin{array}{cccccc}
Q_L &\equiv&({\bf 3, 2,} \frac{1}{3})\subset({\bf 3, 2, 1},\frac{1}{3})
\subset({\bf 4,2,1})\subset{\bf 16}&\equiv& \Gamma\,,\\
L_L &\equiv&({\bf 1, 2,} -1 )\subset({\bf 1, 2, 1},-1)
\subset({\bf 4,2,1})\subset{\bf 16}&\equiv& \Gamma\,,\\
(Q_R)^c_u&\equiv&({\bf \overline{3}, 1}, \frac{2}{3})
\subset({\bf \overline{3}, 1, 2},-\frac{1}{3})
\subset({\bf \overline{4},1,2})\subset{\bf 16}&\equiv&\Gamma\,,\\
(Q_R)^c_d&\equiv&({\bf \overline{3}, 1}, -\frac{4}{3})
\subset({\bf \overline{3}, 1, 2},-\frac{1}{3})
\subset({\bf \overline{4},1,2})\subset{\bf 16}&\equiv&
\Gamma\,,\\
(e_R)^c &\equiv&({\bf 1, 1,} -2 )\subset({\bf 1, 1, 2},1)
\subset({\bf \overline{4},2,1})\subset{\bf 16}&\equiv& 
\Gamma\,,\\
(\nu_R)^c &\equiv&({\bf 1, 1,} 0 )\subset({\bf 1, 1, 2},1)
\subset({\bf \overline{4},2,1})\subset{\bf 16}&\equiv& 
\Gamma\,.\\
\end{array}
\end{eqnarray}
Similarly the various Higgs fields transform as:
\begin{eqnarray}
\begin{array}{cccccc}
T^*_L&\equiv&({\bf \overline{3}, 2,}- \frac{1}{3})\subset(\bf{\overline{3}, 2,
1,} {-\frac{1}{3}})
\subset({\bf \overline{4},2,1})\subset{\bf \overline{16}}&\equiv& 
\Gamma^\dagger\,,\\
h^*_L&\equiv&({\bf 1, 2,} 1 )\subset({\bf 1, 2, 1},1)
\subset({\bf \overline{4},2,1})\subset{\bf \overline{16}}&\equiv& 
\Gamma^\dagger\,,\\
T^{(-)}_R&\equiv&({\bf 3, 1}, -\frac{2}{3})
\subset({\bf 3, 1, 2},\frac{1}{3})
\subset({\bf 4, 1, 2})\subset{\bf \overline{16}}&\equiv&
\Gamma^\dagger\,,\\
T^{(+)}_R&\equiv&({\bf 3, 1}, \frac{4}{3})
\subset({\bf 3, 1, 2},\frac{1}{3})
\subset({\bf 4,1,2})\subset{\bf \overline{16}}&\equiv&
\Gamma^\dagger\,,\\
\phi^{(+)}_{*R}&\equiv&({\bf 1, 1,} 2 )\subset({\bf 1, 1, 2}, -1)
\subset({\bf 4,2,1})\subset{\bf \overline{16}}&\equiv& 
\Gamma^\dagger\,,\\
\phi^{(0)}_{*R}&\equiv&({\bf 1, 1,} 0 )\subset({\bf 1, 1, 2}, -1)
\subset({\bf 4,2,1})\subset{\bf \overline{16}}&\equiv& 
\Gamma^\dagger\,.\\
\end{array}
\end{eqnarray}
In the above expressions the first triplet $({\bf x, y,} z)$
designates the transformation properties under ${\rm G}_{321}$, the
next four numbers $({\bf x, y, z}, w)$ designates the ${\rm G}_{3221}$
transformation properties and the triplet $(\bf{x, y, z})$ stands for
the ${\rm G}_{422}$ transformation properties.

Now we give an order of magnitude estimation about the running of the
effective top Yukawa coupling in our theory. From Eq.~(\ref{mup}) it
is seen that the effective top quark Yukawa coupling looks like
$(\lambda'_\Gamma  v_R f^2)/ (8 \pi^2 M_X)$. In the ${\rm
G}_{321}$ phase it looks like $(\lambda'_\Gamma v_R y_{(qL)}^2)/ (8 \pi^2 M_X)$ 
in the convention adopted to name the Yukawa
couplings in Eq.~(\ref{321}).  Calling this effective coupling as
$y_t$ it will evolve simply like:
\begin{eqnarray}
  {d y_t \over d t }  &=& \frac{y^3_t}{(16 \pi^2)}\,,
\end{eqnarray}
as in the standard-model up to the $M_R$ scale. The gauge and quartic
coupling contributions will be negligible compared to $y_t$. Starting
from the top-quark mass at the electroweak scale, the evolution
equation gives the effective top-quark Yukawa coupling at the
left-right symmetry breaking scale to be of the order of $0.76$ for
$M_R \sim 10^{11}$ GeV. Since the effective coupling constant is a
product of three couplings $(\lambda'_\Gamma v_R y_{(qL)}^2)/ (8 \pi^2 M_X)$
and if $v_R \sim M_X$ then each of these
couplings can individually take values large enough as $3.9$.  As a
result the Yukawa sector becomes non-perturbative in the ${\rm
G}_{3221}$ phase. But on the other hand if we have $(M_X/v_R)\sim
10^{-2}$ then the situation changes. In this case the individual
coupling becomes of the order of $0.84$ which implies the theory is
still perturbative. $M_X$ can be the heavy gauge boson masses or the
singlet fermion $S_L$ mass. The previous condition means that the
heavy gauge bosons or the singlet fermion cannot be heavier than
$10^{9}$ GeV in our theory if we take $M_R \sim 10^{11}$
GeV for a perturbative scenario in the Yukawa sector.

Above the left-right symmetry breaking scale $M_R$ up to the
unification scale $M_U$, the coupling constants $\Lambda'_\Gamma$ and
$f$ will evolve separately. The separate Yukawa couplings remains
finite up to the unification scale.
%
%
\section{Leptogenesis}
\label{lepto}
Since the neutrino masses now depend on the couplings with the
singlets, there is no stringent restriction coming from the up
quark masses. As a result, it may be possible to get large
neutrino mixing angles. The right-handed neutrinos and the new
singlet fermions can now decay into light leptons. The Majorana
masses of the left-handed and right-handed singlets violate lepton
numbers, which in turn can generate enough lepton asymmetry.
Before the electroweak phase transition this asymmetry can then
generate a baryon asymmetry of the universe \cite{fy}. Since there
is no supersymmetry, the gravitino bounds are not present. The
out-of-equilibrium condition can be satisfied near the GUT scale
since the couplings are large to get the required neutrino mass
with large see-saw scale. In this model there is another
interesting feature that the singlets combine with the
right-handed neutrinos to form pseudo-Dirac particles and hence
resonant leptogenesis may also be possible \cite{rlepto1,rlepto2}.

For leptogenesis, consider the interactions of Eq.~(\ref{yukawac}). 
Unlike usual see-saw models with triplet Higgs scalars
\cite{trip,lrlepto}, in this model the right-handed neutrinos cannot decay
into left-handed neutrinos and Higgs bi-doublets dominantly. The
simplest lepton number violating interactions come from the decays
of $S_i$:
\begin{eqnarray}
S_{i}  &\to& \ell_{jL} + \chi^*_L\,, \nonumber \\
   &\to&  {\ell_{jL}}^c + {\chi_L} \,.\label{Si}
\end{eqnarray}
The Majorana masses of $S_i$ allow the singlet to decay into both
leptons and antileptons violating lepton number by two units. In
the present model both $\ell_L$ and $\chi_L$ are very light and
hence these decays are allowed.

\begin{figure}[hb!]
\begin{center}
\epsfxsize12cm\epsffile{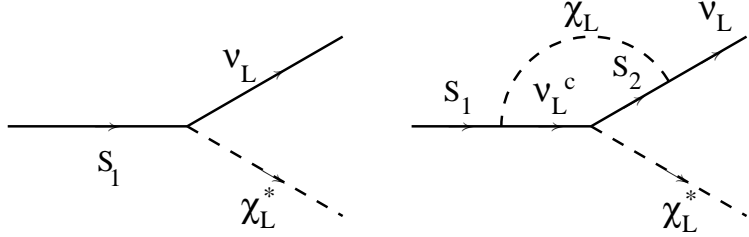} \caption{Vertex type diagrams
interfering with tree level diagram. This is similar to the CP
violation coming from the penguin diagram in K--decays.}
\label{vertex}
\end{center}
\end{figure}

\begin{figure}[hb!]
\begin{center}
\epsfxsize12cm\epsffile{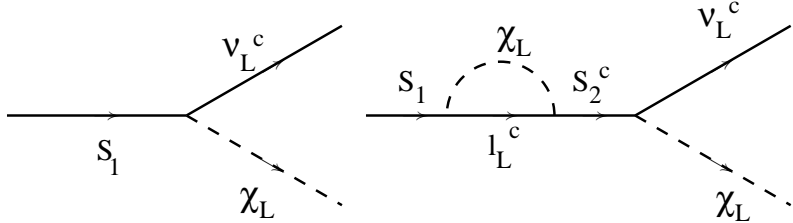} \caption{Self energy diagram
interfering with tree level diagram. This is similar to CP
violation in $K-\bar{K}$ oscillation, entering in mass matrix of
$S_i$.} \label{selfenergy}
\end{center}
\end{figure}

For CP violation there are two types of one-loop diagrams which
interferes with the tree-level diagrams for the decays of $S_i$.
These are the vertex type diagrams of Fig.~\ref{vertex} and
Fig.~\ref{selfenergy}.

The right-handed neutrinos do not take part in leptogenesis
directly, but due to mixing of the right-handed neutrinos with the
heavy singlets $S_i$, the right-handed neutrino decays also enter
in the picture of leptogenesis. In the limit of $M_S \gg m_R$, the
right-handed neutrinos and the $S_i$ are both heavy and distinct.
In this case the amount lepton asymmetry due to  CP violation is given by,
%
%
%
\begin{equation}
\delta  = - {1 \over 8 \pi} \frac{M_1 }{M_2} \frac{{\rm Im} [
\sum_\alpha (k'^\ast_{(l)\,\alpha 1} k'_{(l)\,\alpha 2})
 \sum_\beta (k'^\ast_{(l)\,\beta 1} k'_{(l)\,\beta 2}) ] }{ \sum_\alpha
| k'_{(l)\,\alpha 1}|^2}\,,
\end{equation}
where we assumed that the mass matrix of $S_i$ are diagonal and
the eigenvalues are hierarchical $M_1 < M_2 < M_3$ where $M_i$ are
masses of $S_i$. We can write the effective lepton asymmetry as:
\begin{eqnarray}
\epsilon = \delta \zeta\,,
\label{la}
\end{eqnarray}
where $\zeta$ is a suppression factor which depends on the amount of 
departure from equilibrium. The exact value of $\zeta$ can be
obtained by solving the Boltzmann equation taking all the interactions
into consideration. However, it is also possible to make an order
of magnitude estimate for the amount of asymmetry, which will be
very close to the actual value. 
Since $S_1$ is the lightest of the singlets, the
decay of this singlet will be able to generate the lepton
asymmetry. The asymmetry generated or washed out by the heavier
ones $S_2$ or $S_3$ will be smeared out by the interactions of
$S_1$ after $S_2$ and $S_3$ had decayed away. So, for an estimate
we shall only consider the decays of $S_1$. This assumption is
well justified when the singlets $S_i$ have a hierarchical mass
structure.

The out-of-equilibrium condition is parametrized by:
\begin{equation}
\theta=\frac{1.7\sqrt{g_*} {T^2 \over M_P}}
{{|k'_{(l)\,\alpha 1}|^2 \over 16 \pi} M_1}\,, \hskip .1cm {\rm at}\quad T = M_1\,.
\end{equation}
When $\theta > 1$, there is no Boltzmann suppression of the
generated asymmetry and the out-of-equilibrium condition is
satisfied. In this case the generated asymmetry is given by 
$\delta$ and it is not
washed out after it is created and one gets $\zeta = 1$. However,
if $\theta \gg 1$, then the interaction strength is so slow 
that the generated asymmetry can never reach the value $\delta$.
Although the interactions cannot wash out the asymmetry after
it is generated, the amount of asymmetry is less than $\delta$. 
In the case of $\theta < 1$, the generated asymmetry is same
as the CP asymmetry of $\delta$, but even after the asymmetry
is created, the interaction remains strong enough to deplete
the asymmetry. Although the depletion is exponentially fast,
it cannot compete with the expansion of the universe for long
and the final amount of asymmetry is not exponentially depleted.
It was shown that \cite{lola} the suppression factor $\zeta$ is almost
linearly proportional to $\theta$. In the present model we come
across this last scenario.

In the present model $10^9 ~ {\rm GeV} < M_R < 10^{11}$ GeV.
While a lower value of $M_R$ is preferable for out-of-equilibrium
condition, since the Yukawa couplings grow very fast above the
scale $M_R$ we have to consider the highest value of $M_R$. 
Taking the hierarchical structure of $S_i$, we consider the
mass of the lightest singlet to be around $10^{10}$ GeV.
Taking $M_1 \sim 10^{10}$ GeV, $g_*\sim 10^2$ and 
$ k'_{(l)\,\alpha 1}\sim 1$ we find that $\theta$ is much
lower than 1, which gives a strong suppression factor
of $\zeta \sim 10^{-7}$. On the other hand the Yukawa
couplings in this model comes out to be of the order
of 1 and hence we get a large enhancement in the CP asymmetry
and $\delta$ in our case can be as large as $10^{-2}$ and 
so the lepton asymmetry parameter $\epsilon \sim 10^{-9}$. 

At this stage, $\Delta B = 0$ and $ \Delta L$ is given by
$\delta$. Thus 
\begin{equation}  \Delta (B-L) = \epsilon\,.\end{equation}
The final baryon
asymmetry after the electroweak phase transition is thus given by,
\begin{equation}  {n_B \over s} = {1 \over 3} \Delta (B-L) = {\epsilon \over 3}
\sim 10^{-10}\,.
\end{equation}

In the case of $M_S \ll m_R$, the right-handed neutrinos and the $S_i$
singlets of every generation are almost degenerate. The mass splitting
between the states $\nu_{i R}$ and $S_i$ with mass $m_R$ is of the
order of $M_S$. Although $S_i$ decays will now generate a lepton
asymmetry, both the heavy mass eigenstates contain the states
$S_i$. As a result, when these two almost degenerate states decay,
there may be resonant leptogenesis (which will however require new
interactions and fine-tuning) and hence the scale of
leptogenesis could be very low.  For the present scenario since $M_R$
and hence $M_1$ cannot be much lower, this is not important and hence
we shall not discuss it in any further detail.
\section{Summary}
In conclusion, we constructed an SO(10) GUT without any Higgs
bi-doublets. All the symmetry breaking could be achieved by only two
Higgs scalars, a ${\bf 210}$ and a ${\bf 16}$. By including a massive
singlet fermion per generation we break chiral symmetry which can then
give masses to all the fermions radiatively without introducing any
new scalar fields. All fermion masses have the same see-saw form. The
spontaneous parity breaking plays a crucial role in breaking the
left-handed and right-handed SU(2) groups at two widely different
scales and also giving masses to the left-handed neutrinos in this
scenario.  The spontaneous breaking of an ungauged discrete
symmetry, the D-parity, which is a special feature of this model may 
cause formation of very heavy domain walls of GUT-scale mass. 
Since the GUT scale in the model is rather low $(\sim 10^{15}$
GeV), some of the parameters of the model may have to be constrained 
further to prevent fast proton decay. These features require further study and 
will be taken up in future. The model allows large neutrino mixing and required
neutrino masses. The baryon asymmetry of the universe can be
explained through leptogenesis.

\section*{Acknowledgement} GR acknowledges the support of the 
Raja Ramanna Fellowship of the Department of Atomic Energy, 
Government of India and the hospitality of the Physics
Department, University of California, Riverside. BRD acknowledges
the support in part by the U.S. Department of Energy under Grant
No. DE-FG03-94ER40837.

\clearpage\newpage

\end{document}